# From LIGO to Fiber: Compact Sagnac Interferometer for Gravitational-Wave Detection

(Dated, November 4th, 2025)

Farhad Hakimi and Hosain Hakimi, emails: fhakimi@alum.mit.edu, hhakini@alum.mit.edu

**Abstract**

Gravitational-wave (GW) detection has transformed astrophysics, granting us direct access to black hole mergers, neutron star collisions, and the cataclysms of stellar death. Yet the great observatories of today—LIGO, Virgo, KAGRA, and the planned Einstein Telescope—rest on Michelson interferometers that, despite their triumphs, confront fundamental barriers of scale, cost, and environmental vulnerability.
We envision a new path: a Sagnac-based fiber interferometer that leverages reciprocity and inherent robustness. Its meter-scale, modular design—compact enough to fit within a small facility—offers dramatic gains in scalability and affordability over kilometer-scale Michelson systems. Tunable to frequency bands where conventional detectors lose sensitivity, it opens the door to compact, versatile, and accessible GW observatories, empowering universities and research centers worldwide. Linked together in a global network, such facilities could transcend mere detection: they could localize cosmic sources and reconstruct them into images—much as black holes were first directly revealed—ushering in a new era of gravitational-wave astronomy and multi-messenger discovery.

**I. Introduction:**

Laser interferometers use laser beams to detect incredibly small distance changes caused by gravitational waves, ripples in space-time predicted by Einstein. LIGO, built in the 1990s with sites in Louisiana and Washington, became fully operational in 2015 and soon detected the first gravitational wave, GW150914, which shifted its 4-km arms by just one-thousandth the diameter of a proton [1]. Now, the U.S. is developing the Cosmic Explorer, a next-generation interferometer with 40-km arms—ten times longer than LIGO's—to achieve even greater sensitivity [2].

The quest to build bigger detectors echoes a familiar story. In the 1700s, sailors tried to solve the problem of finding longitude by making clocks larger and more elaborate—until John Harrison changed everything with his small, brilliantly engineered marine chronometer. The answer wasn't size, but design. Likewise, the future of gravitational wave astronomy may rest not only on scaling up, but on reimagining the approach—crafting technologies that are as practical as they are powerful.

Remarkably, advanced interferometric fiber-optic gyroscopes (IFOGs) can already detect path-length differences on the order of a proton's diameter [3–4], using only a few kilometers of fiber wound into a coil just a few inches in diameter. With targeted modifications that integrate existing technologies, a redesigned IFOG has the potential to rival the sensitivity of large ground-based gravitational wave detectors.

The rise of megahertz-speed fiber-optic communications in the 1970s unexpectedly laid the groundwork for fiber-optic gyroscopes. Today, history is on the verge of repeating itself: the advent of ultra-high-speed (terahertz-range), yet compact, communication technologies is enabling a new generation of ultra-sensitive optical sensors—heralding a transformative era for gravitational wave (GW) detection.

In the 1970s, two main fiber-optic gyroscope designs emerged: the interferometric fiber-optic gyroscope (IFOG) and the resonant fiber-optic gyroscope (RFOG). IFOGs use ~1 km fiber loops with single-pass, counter-propagating beams to detect rotation via interferometric fringe shifts, while RFOGs employ much shorter loops (~10 m) with light circulating multiple times (~100 turns) to build sensitivity. IFOGs operate with broadband sources like LEDs, whereas RFOGs require narrow-linewidth lasers with long coherence lengths. However, RFOGs are prone to errors from fiber imperfections—backscattering, birefringence, polarization coupling, and nonlinearities—largely suppressed in IFOGs, which consistently outperform them.

The key insight is that many fiber-sensor errors can be suppressed by using a broadband source within a balanced interferometric setup—most notably, a Sagnac configuration. The critical requirement is reciprocity: the clockwise and counterclockwise beams must traverse identical paths so that any residual phase shift arises solely from rotation. This principle underpins the "minimum-configuration" IFOG (Fig. 1 [3–4]), which achieves near-perfect reciprocity and minimal error. It is this architecture that forms the basis of the fiber-optic implementation proposed for the GW detector in this work.

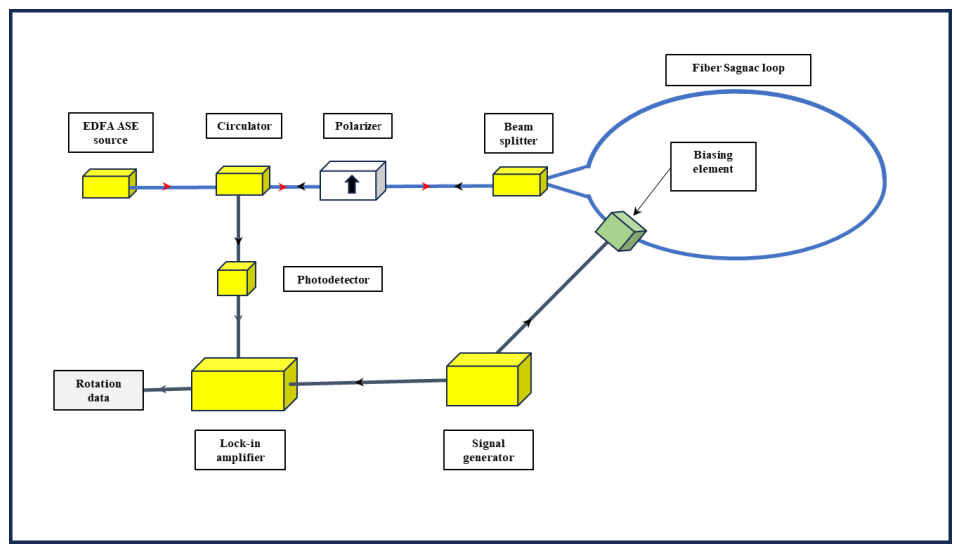

**Figure 1**: Interferometric Fiber Optic Gyroscope (IFOG) minimum-configuration architecture.

**II. The Proposed New Gravitational Wave Detector:**

In 1996, Sun *et al.* at Stanford University [5] demonstrated that free-space Sagnac interferometers can offer advantages over conventional Michelson configurations in certain regimes. The Stanford studies highlighted key benefits of the Sagnac configuration, including intrinsic common-mode noise rejection and greater geometric flexibility in the optical layout [5-6]. Their design adopted a free-space geometry distinct from LIGO's Michelson interferometer, while still relying on similar core components such as bulk optics, coherent laser sources, and sensitive photodetectors. Achieving competitive sensitivity required kilometer-scale arms comparable to LIGO's 4-km baseline, with some custom optical components to accommodate the new geometry.

Nevertheless, practical implementation faced significant obstacles. The specialized optics demanded by the Sagnac design imposed high costs and technical complexity. Combined with the rapid progress and demonstrated success of Michelson-based observatories like LIGO, these factors led the community to favor more mature and cost-effective approaches, leaving Sagnac-based concepts largely unexplored.

However, the Stanford design overlooked a transformative opportunity: replacing kilometer-scale free-space structures and narrowband lasers with long optical fibers powered by a broadband source—such as erbium-doped fiber amplifiers amplified spontaneous emission (EDFA-ASE)—that inherently suppresses most fiber-related impairments. This shift not only streamlines the interferometer architecture but also eliminates the dominant sources of degradation that have historically constrained performance. Building on this insight, the present work introduces a compact Sagnac geometry that leverages the proven precision and robustness of state-of-the-art fiber-optic gyroscopes—delivering high sensitivity within a dramatically smaller and more practical footprint.

Inspired by the exceptional sensitivity of IFOGs and their immunity against many fiber-related limitations, the proposed gravitational wave (GW) Sagnac sensor adopts a similar geometric configuration (Fig. 2). The sensor's fiber loop is configured in an L-shape to match the time-varying strain of GWs, specifically the h+ polarization mode propagating along the z-axis (Fig. 3).

While rooted in the conventional IFOG architecture, this GW detection sensor integrates several enhancements to elevate performance: a relative intensity noise (RIN) suppressor, a high-speed intensity modulator, high-speed photodetectors, and a high-speed analog-to-digital converter (ADC) with an advanced data post-processor for waveform extraction.

A key feature of the system is the phase bias within the Sagnac loop. Although the bias can be implemented either actively or passively, the passive method is preferred here for its simplicity and stability. To minimize interferometer drift, the light must remain fully depolarized throughout the loop.

Unlike LIGO, which suppresses phase noise using techniques such as squeezed light, this design targets intensity noise suppression—providing a distinct yet complementary path to improved sensitivity.

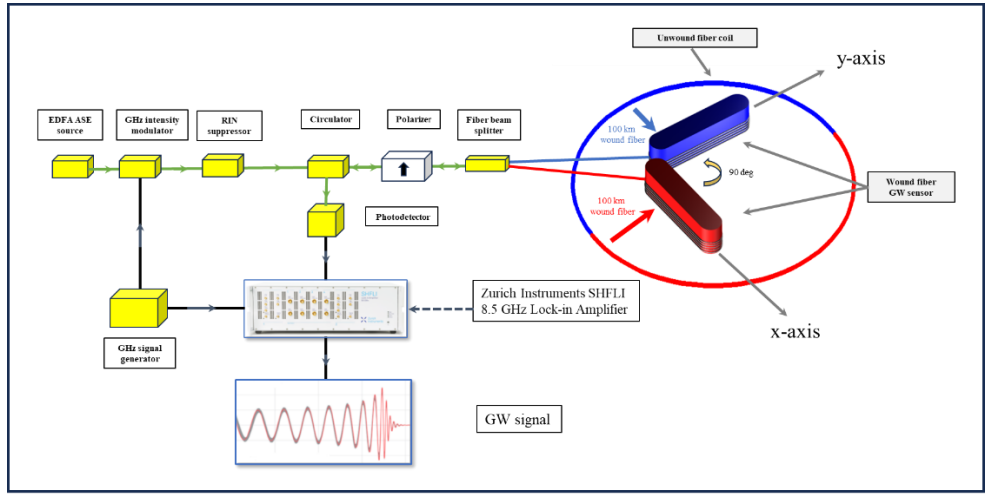

Figure 2: General architecture of the L-shape Sagnac gravitational wave detector.

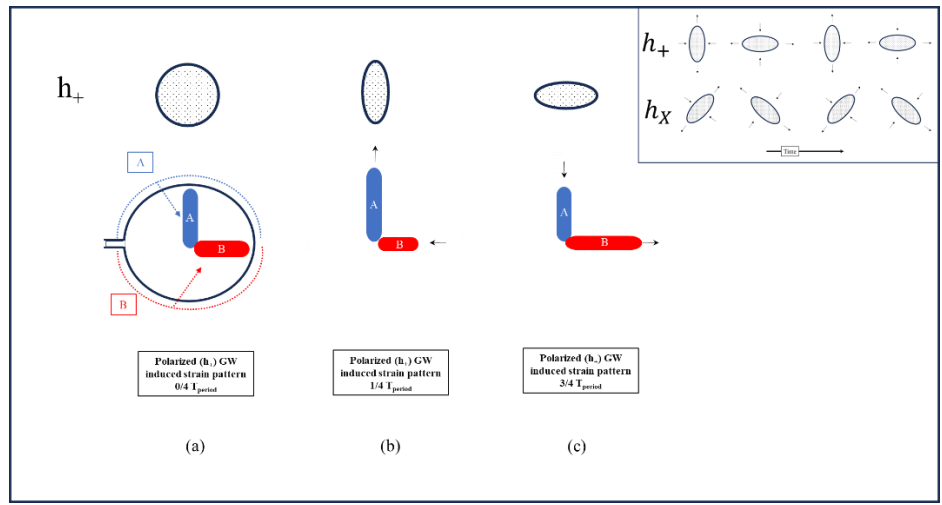

**Figure 3**: a) Each half of the unwound Sagnac loop is wound into the corresponding lobe, as shown, to match the evolution of the h⁺ gravitational wave (GW) strain for maximum sensitivity. b) and c) illustrate the stretching and contracting of the L-shape sensor in response to an h⁺ GW propagating perpendicular to the page. The inset shows the time-varying strain patterns induced by h⁺ and h× polarizations of a GW on a circular object, for a wave propagating out of the page.

The fiber Sagnac sensor is configured in an L-shape to match the strain pattern induced by a passing gravitational wave (GW). Fig. 3a shows both the unwound and L-shape configurations. To achieve rotational insensitivity, the sensor's net enclosed area must be zero, which is ensured by winding the horizontal and vertical arms in opposite directions.

Fig. 3b and 3c depict a gravitational wave (GW) with h+ polarization propagating perpendicular to the page (along the z-axis), distorting the L-shape sensor. During the first half of the GW cycle, the sensor stretches along the y-axis and contracts along the x-axis; in the second half, the effect reverses, with stretching along x and contraction along y. This alternating strain induces a phase shift in the Sagnac sensor during the first half-cycle, followed by an equal and opposite shift during the second half (Fig.4).

**III. Sagnac Fiber Gravitational Wave Detector Performance (signal to noise):** Assessing the sensitivity of a fiber-based Sagnac GW detector demands a rigorous comparison between the expected signal response and the cumulative noise contributions from optical components, fiber propagation, and photodetectors. The resulting signal-to-noise ratio (SNR) can then be quantified as outlined below.

**A. GW Signal Strength:**

An interferometer biased at the quadrature point, as shown in Fig. 4, generates a gravitational-wave (GW) signal of frequency $f_{GW}$ according to Eqs. (1)–(6). Here, $P_{opt}$ is the total optical power received by all detectors, $h$ is the GW-induced strain, $S$ is the scale factor, $n$ is the effective refractive index of the fiber, $L_{fiber}$ is the fiber length, and $\lambda_{light}$ is the average wavelength of the broadband light source. The biasing element is set to 90° (the quadrature point), placing the interferometer at its point of maximum slope for optimal detection sensitivity (Fig. 4).

$$P = P_{opt}(1 + \cos(\phi_{Bias} + \phi_{GW})); \tag{1}$$

$$\phi_{GW} = S\,h\,\sin(2\pi f_{GW} t); \tag{2}$$

$$\phi_{Bias} = \frac{\pi}{2}; \tag{3}$$

$$S = 2\pi n \left(\frac{L_{fiber}}{\lambda_{light}}\right). \tag{4}$$

Expanding the cosine term in Eq. (1) and applying the Jacobi–Anger expansion yields

$$P \approx P_{opt}[1 - 2J_1(Sh)\sin(2\pi f_{GW} t)]. \tag{5}$$

where $J_1$ is the Bessel function of the first kind. The second sinusoidal term constitutes the GW signal. For small values of $S\,h$, the signal term simplifies to

$$P_{GW-signal} = -2\,P_{opt}\,J_1(S\,h)\,\sin(2\pi f_{GW} t) \approx -P_{opt}\,S\,h\,\sin(2\pi f_{GW} t). \tag{6}$$

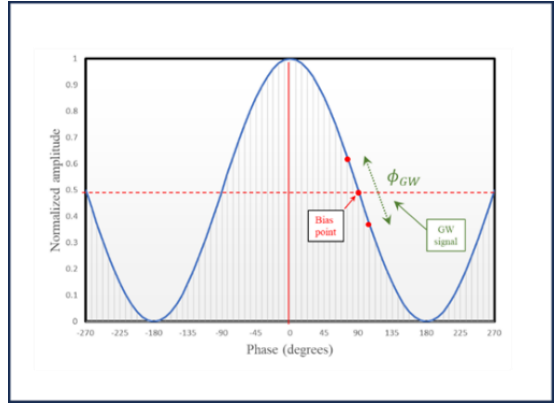

**Figure 4:** The biasing element sets the interferometer at quadrature—its point of maximum slope—to achieve optimal sensitivity.

**B. Noise in Optical Source:** In a Sagnac interferometer with broadband optical sources, one of the primary performance limits is relative intensity noise (RIN)—random fluctuations in optical power that can mask signals of interest, such as distortions caused by a passing gravitational wave. Unless suppressed, RIN generally dominates over optical shot noise.

A practical suppression method uses a two-stage optical scheme. First, a broadband source is generated, typically from amplified spontaneous emission (ASE) in an erbium-doped fiber amplifier (EDFA). ASE provides a wide spectral range with relatively uniform power but suffers from high RIN. To mitigate this, the ASE output is passed through a semiconductor optical amplifier (SOA) operated in the deep saturation regime, where the high input power nearly depletes the SOA's gain. In this state, the SOA acts as a power equalizer, effectively flattening intensity fluctuations.

Commercial SOAs operated in this regime have demonstrated up to 22 dB of RIN suppression, corresponding to a hundredfold reduction in noise power [7]. The principal limitation of this method is bandwidth: RIN suppression is typically effective only up to ~4 GHz, thereby restricting the portion of the noise spectrum that can be mitigated.

**C. Noises in Fiber:** Insights from the development of fiber-optic gyroscopes can be directly applied to gravitational-wave fiber detectors—particularly the use of broadband optical sources with short coherence lengths, which help mitigate error sources such as self-phase modulation and cross-phase modulation among others. Nevertheless, two intrinsic fiber noise sources remain significant: *thermo-optic noise,* arising from temperature-induced variations in refractive index and fiber length; and *thermomechanical noise,* longitudinal fluctuations caused by random thermal vibrations along the fiber due to mechanical dissipation. These vibrations produce minute length changes (longitudinal strain), altering the optical path length and introducing phase noise in interferometric systems. Thermomechanical noise, however, is suppressed in balanced interferometers such as the Sagnac, owing to their inherent common-mode rejection.

K. H. Wanser [8], in his seminal paper *"Fundamental Phase Noise Limit in Optical Fibers Due to Temperature Fluctuations,"* derived the governing expressions for these noise sources in long optical fibers. His analysis shows that both thermo-optic and thermomechanical noise decrease markedly at higher frequencies and lower temperatures. This makes it essential to operate a Sagnac-based gravitational-wave detector at the highest practical frequencies and the lowest achievable temperatures.

Figure 5 shows the frequency dependence of thermo-optic noise at three temperatures, highlighting the combined benefits of high-frequency operation and cryogenic cooling. As a representative case, a 200 km single-mode fiber cooled to the superfluid helium regime (~2 K) exhibits phase noise on the order of –100 dB/√Hz near 1.8 GHz—substantially lower than at room temperature or in the low-frequency band. Moreover, the exceptionally high effective thermal conductivity of superfluid helium ensures near-uniform temperature along the entire fiber, thereby suppressing longitudinal thermal gradients and associated refractive-index fluctuations.

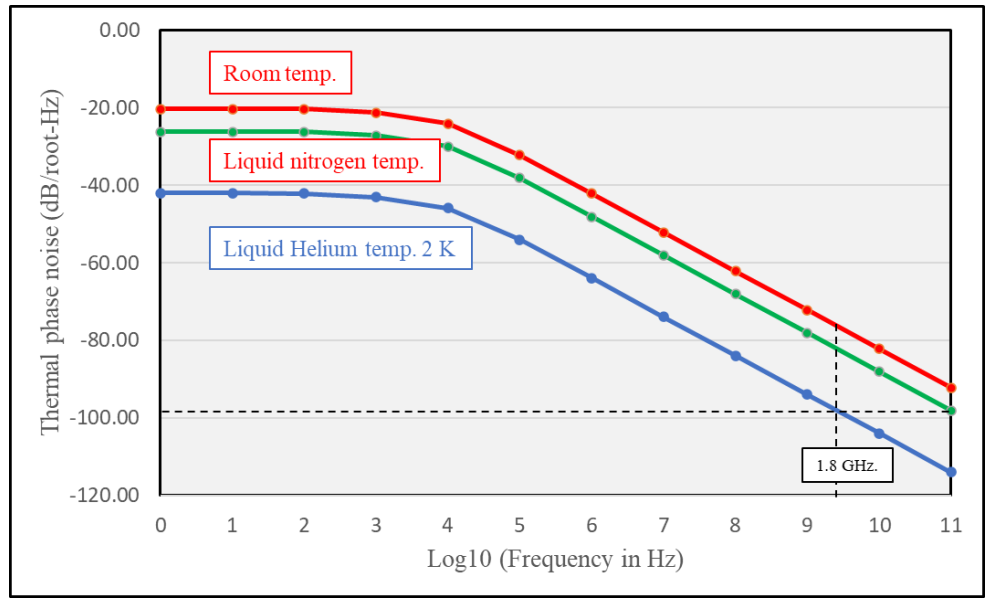

**Figure 5**: Phase noise of a 200 km single-mode fiber as a function of frequency at three different temperatures. The plot highlights the advantage of operating at higher frequencies and lower temperatures to reduce phase noise.

**D. Noise sources in the detection system:** The detection system is limited by three fundamental noise contributions: optical shot noise, optical RIN noise, and electronic noise. In the following definitions, $h\nu$ denotes the photon energy, $P_{opt}$ is the total optical power incident on all detectors, $\lambda_{light}$ is the mean optical wavelength, NEP is the detector noise-equivalent power, $R$ is the RIN suppression factor, $BW_{elec}$ is the electronic bandwidth, and $\Delta\nu_{optical}$ is the optical bandwidth. Expressions (7)–(9) quantify the respective contributions from each noise mechanism.

$$\text{Shot noise} = \sqrt{\frac{2hc}{\lambda_{light}} P_{opt} \, BW_{elec}} \, ; \quad (7)$$

$$RIN_{opt} = \frac{1}{R} P_{opt} \sqrt{\frac{BW_{elec.}}{\Delta v_{opt}}} \, ; \quad (8)$$

$$\text{Elec noise} = NEP \sqrt{BW_{elec}} \, . \quad (9)$$

The equations above define the strain-equivalent gravitational-wave (GW) noise contributions—shot noise, electronic noise, and relative intensity noise (RIN)—per unit bandwidth, as summarized below:

$$h_{shot} = \frac{1}{S} \sqrt{\frac{2hc}{\lambda_{light} P_{opt}}} \, ; \quad (10)$$

$$h_{RIN} = \frac{1}{S} \frac{1}{R} \frac{1}{\sqrt{\Delta v_{opt}}} \, . \quad (11)$$

$$h_{elec} = \frac{1}{S} \frac{NEP}{P_{Opt}} \, ; \quad (12)$$

**IV. A Example Implementation 1:** Let us consider an example of an L-shaped Sagnac detector optimized for detecting 1 kHz gravitational waves, a frequency comparable to that of LIGO. To illustrate the design, we assign representative parameters to the proposed configuration. A 200 km single-mode fiber provides a light transit time of approximately 1 millisecond. The biasing element is set to 90°, positioning the interferometer at its point of maximum slope to maximize detection sensitivity, as shown in Fig. 4.

The amplified spontaneous emission (ASE) depolarized light from the erbium-doped fiber amplifier (EDFA) is intensity-modulated at 1.8 GHz, below the relative intensity noise (RIN) suppression bandwidth of the semiconductor optical amplifier (SOA). The modulated signal is then injected into the SOA to suppress RIN. Assuming 300 mW [9] of input power is launched into the 200 km fiber sensor, and the fiber attenuation is 0.15 dB/km, approximately 0.3 mW reaches to a high-speed photodiode.

Figure 6 illustrates the impact of different noise sources on a gravitational-wave Sagnac detector. The plot shows that relative intensity noise (RIN) dominates the sensitivity limit unless it is strongly suppressed. With effective RIN reduction, the detector can reach its targeted sensitivity.

The values to generate Fig. 6 plot are, wavelength 1.5 μm, optical bandwidth of 30 nm ($3.75 \times 10^{12}$ Hz), 14 dB RIN-suppression, electronic bandwidth of 1 kHz, and 0.3 mW optical received by the high-speed photo-detector with NEP of $2 \times 10^{-12}$ W/√Hz. The figure shows the strain detection sensitivity h of approximately 10$^{-18}$ is possible before post-data processing with a single stage Sagnac interferometer. To put this in context, first generation LIGO had strain detection sensitivity of 3x10$^{-20}$ at 1 kHz frequency.

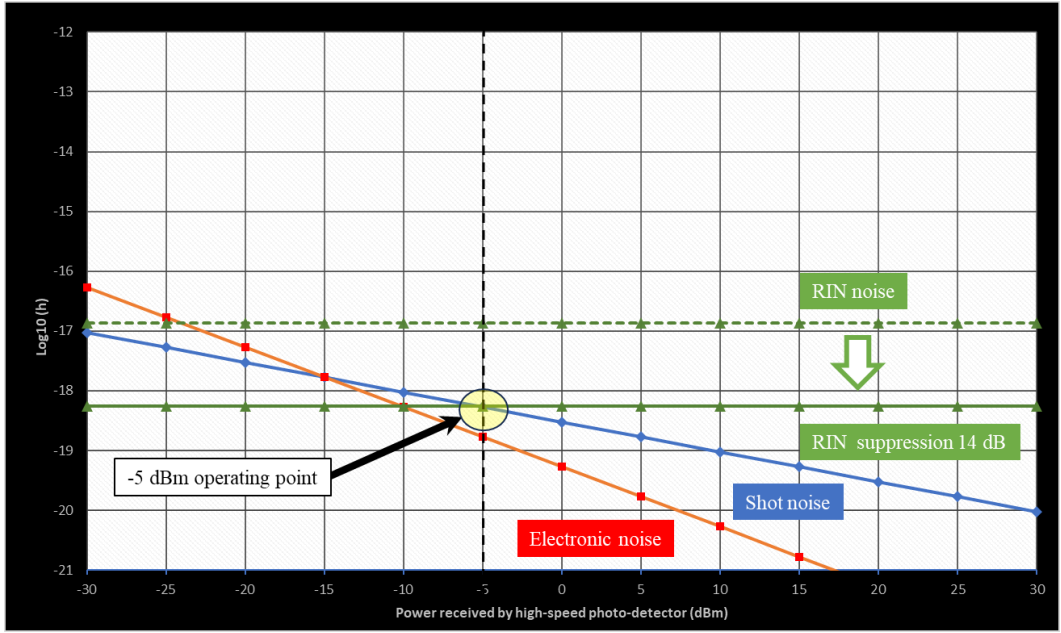

**Figure 6:** Illustrates the various noise contributions affecting sensor sensitivity, including relative intensity noise (RIN) before and after suppression. A raw (pre-processed) strain sensitivity of approximately $h \sim 10^{-18}$ can be achieved at a received power of -5 dBm by a photodiode. For comparison, the first-generation LIGO achieved a strain sensitivity of $3 \times 10^{-20}$ at 1 kHz.

Gravitational-wave (GW) signals reside at low frequencies, while the Sagnac detector waveform is modulated at 1.8 GHz. As a result, the received signal must be demodulated using a demodulator to recover the GW signal at baseband. The fidelity of this demodulated signal can be dramatically enhanced through digital post-processing with oversampling, enabling a significant increase in the effective signal-to-noise ratio (SNR). The achievable SNR improvement is fundamentally determined by the oversampling ratio (OSR) and the effective number of bits (ENOB) of the analog-to-digital converter (ADC). For a commercial 12-bit ADC operating at 10.4 GSa/s [10], the corresponding oversampling ratio is, allowing for SNR gains that push the limits of GW detection sensitivity.

$$\text{OSR} = \frac{10.4 \times 10^9}{2 \times 10^3} = 5.2 \times 10^6. \tag{13}$$

The factor of two in the denominator comes from the Nyquist sampling criterion. This yields an ideal oversampling gain of approximately 10 log$_{10}$(OSR) ≈ 67 dB—an OSR-limited improvement in SNR.

The quantization-limited SNR of a 12-bit ADC is given by

$$\text{SNR}_{\text{ADC}} = 6.02 \times 12 + 1.76 \approx 74 \text{ dB}. \tag{14}$$

Accordingly, the theoretical upper limit to the overall improvement is

$$\min(74 \text{ dB}, 67 \text{ dB}) = 67 \text{ dB}. \tag{15}$$

In real-world systems, the ideal SNR gain is inevitably degraded by ADC thermal noise, sampling jitter, and nonlinearities, limiting the achievable improvement to roughly 50 dB. Even so, the raw (pre-processed) strain sensitivity shown in Fig. 6—$h \approx 10^{-18}$ over a 1 kHz bandwidth—could still be boosted by about 50 dB through digital post-processing, driving the effective sensitivity down to $h \approx 10^{-23}$. Remarkably, this rivals the performance of Advanced LIGO—but achieved in a single-stage configuration.

Even more compelling, the required technology is not hypothetical. GHz digital lock-in amplifiers with exactly these capabilities are already on the market [11,12]. For instance, Zurich Instruments' 8.5 GHz lock-in amplifier provides a 6 GSa/s digitizer and a 14-bit ADC with 100 dB of dynamic reserve, making it a direct match for the implementation described above.

**B. Example Implementation 2**: In this approach, the system operates at room temperature, but the optical source is modulated at a much higher frequency—around 100 GHz—to exploit the approximately 20 dB lower fiber noise compared to operation at 1.8 GHz. The resulting signal is then down-converted to baseband using an RF mixer and digitized via oversampling, as described earlier. Although this method does not achieve the same sensitivity as the first implementation example, it remains a practical option when cryogenic cooling is not available.

To boost sensitivity, gravitational-wave interferometers could follow the same strategy that transformed astronomy: replacing a single massive mirror with an array of smaller, near-perfect ones—exemplified by JWST, ELT, and TMT. Likewise, multiple Sagnac stages, stacked or aligned, could be made to sense the same wave, and their outputs combined to multiply sensitivity. Even a single-stage Sagnac, as outlined above, already approaches the performance of Advanced LIGO. While Einstein Telescope and Cosmic Explorer target nearly 10× improvement over Advanced LIGO—the Sagnac architecture offers a fundamentally different scaling path. In the simplest case, where outputs are added incoherently, sensitivity scales as the square root of the number of stages, $\sqrt{N}$, due to statistical averaging of uncorrelated noise.

In principle, a detector with just a few hundred stages could rival the Einstein Telescope or Cosmic Explorer. Critically, the physical footprint need not balloon: hundreds to thousands of independent fiber loops can be co-located within a single liquid helium bath, keeping the total cryostat volume well within practical engineering limits. The implication is transformative: an observatory-class gravitational-wave detector—compact, cryogenically stabilized, shielded from environmental noise, and fully scalable—could fit inside a small single research facility while rivaling the most ambitious ground-based observatories.

**V. Discussion:** By reimagining the Interferometric Fiber Optic Gyroscope (IFOG)—already an exquisitely sensitive sensor—it is possible to create gravitational-wave detectors of unprecedented precision. All key enabling technologies are mature, commercially available, and already demonstrated: high-speed modulators and photodetectors, erbium-doped and semiconductor optical amplifiers, low-loss optical fibers, precision electronics, and cryogenic systems operating at 1.9 K (e.g., from Linde Kryotechnik AG and ICEoxford). This means there are no fundamental barriers to realizing a next-generation fiber-based GW observatory capable of complementing—or even rivaling—today's largest facilities.

Furthermore, a second L-shaped Sagnac interferometer placed 45° relative to the first enables direct measurement of GW polarization, providing insight into the orientation of distant binary systems. Measuring polarization also offers a way to test alternative theories of gravity, as discussed later in the article.

Additionally, the fiber lengths in the Sagnac configuration can be adjusted to target specific gravitational-wave frequency bands. By tuning the optical path length, the interferometer's response can be optimized for different signal frequencies, further enhancing the system's versatility and adaptability to a range of astrophysical scenarios.

**VI. Outlook:** Forthcoming breakthroughs in photonics and optoelectronics could fundamentally reshape gravitational-wave (GW) astronomy. Air-core optical fibers, now approaching the ultimate attenuation limit of 0.01 dB/km, may soon support sensor links extending thousands of kilometers, without the need for massive vacuum infrastructure. Higher levels of RIN suppression can be achieved by cascading SOAs, although the bandwidth is still limited to about 4 GHz. At the same time, new approaches are needed to deliver much stronger relative intensity noise suppression across dramatically wider bandwidths—ideally reaching into the hundreds of gigahertz.

Figure 6 indicates that increasing the optical power received by the detector—whether through lower-loss fiber (e.g., future air-core fiber) or a higher-power SOA—will require greater RIN suppression to achieve higher sensor sensitivity.

With radically lower cost and effortless scalability, synchronized Sagnac GW detectors could be deployed across the globe, forming a vast, interconnected observatory. In such a network, these facilities would transcend mere detection: they could pinpoint cosmic sources, reconstruct them into images—much as black holes were first directly revealed—and illuminate the birth of gravitational waves themselves.

Equally profound is the opportunity to measure GW polarization directly. While General Relativity predicts only two polarization modes, alternative theories—including massive gravity and string-inspired models—predict five, six, or more (Fig. 7). Current observatories can only infer polarization indirectly through multi-site correlation and complex post-processing. Next-generation Sagnac detectors, by contrast, could disentangle polarization at the point of detection by deploying multiple units with varied orientations.

The day three Sagnac detectors simultaneously capture two transverse modes and a longitudinal mode will mark an epochal discovery—an unmistakable signal of new physics. Like the Copernican shift or the advent of quantum mechanics, it will demand a revision of Einstein's theory of gravitation and inaugurate a new era in our understanding of the universe [13–15].

In conclusion, the rapid convergence of high-speed fiber communication and ultra-sensitive fiber sensing technologies could herald a seismic shift in gravitational-wave astronomy: the emergence of compact, cryogenically stabilized, fully scalable fiber-based detectors that could operate entirely within a single research facility while matching—and ultimately surpassing—the sensitivity of today's multi-kilometer observatories.

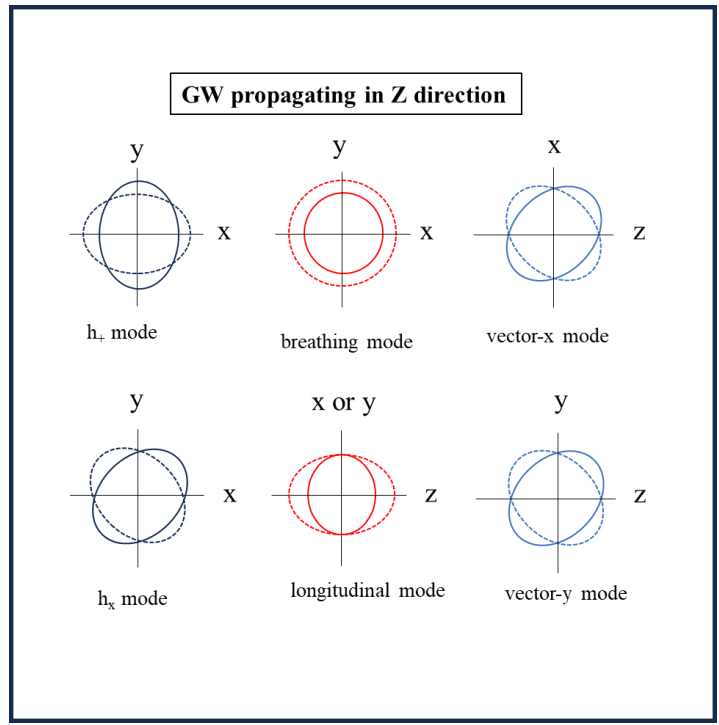

**Figure 7:** Alternative theories of gravity predict polarization states beyond the standard $h_+$ and $h_\times$ modes shown above, and detecting these additional modes requires three mutually orthogonal GW detectors.